\documentstyle{article}

\begin{document}

\topmargin 0pt \oddsidemargin 5mm

\setcounter{page}{1}

\begin{quotation}
\hspace{8cm}Preprint YerPhI-1511(11)-98

\hspace{8cm}{} \vspace{2cm}
\end{quotation}

\begin{center}
{\large {THREE-DIMENSIONAL WAKE FIELDS, GENERATED IN PLASMA \\BY CYLINDRICAL
ELECTRON BUNCH}}\\

S.S. Elbakian, E.V. Sekhpossian, {A.G. Khachatryan}\\ \vspace{1cm} {\em %
Yerevan Physics Institute, Alikhanian Brothers St. 2, Yerevan 375036,
Republic of Armenia}\\ E-mail: khachatr@moon.yerphi.am
\end{center}

\vspace {5mm} \centerline{{\bf{Abstract}}}

The expressions for wake fields, generated in plasma (in the plasma
waveguide or unlimited plasma) by relativistic electron bunch, was received
and analyzed for the cases of the presence and absence of strong external
longitudinal magnetic field. For the both cases the comparative analysis of
the dependence of field amplitudes on the parameters of the electron bunch
was done.

\vspace {5mm}PACS number(s): 52.35.Mw, 52.40.Mj \newpage 

\section{Introduction}

Presently the studies on new methods of charged particle acceleration by
means of wake fields, generated in plasma by laser radiation (BWA (Beat-Wave
Acceleration), LWFA (Laser Wake Field Acceleration)) and by bunches of
relativistic particles (PWFA (Plasma Wake-Field Acceleration)), moving in
plasma are intensively developed (see, e.g. reviews [1, 2] and cited there
literature). The intensity of acceleration fields (in the order of $%
10^7-10^8V/cm$), attained by these methods can be used both for the charge
acceleration, and for focusing of electron (positron) bunches in order to
obtain the beams of high density and to ensure high luminosity in linear
colliders of next generation [1, 3].

Linear theory of wake field generating by two- and three-dimensional rigid
bunches of charged particles in boundless and limited plasma was developed
in many works [3-11]. Nonlinear theory of wake field generating by a rigid
one-dimensional bunch of final extent and sequence of charge particle bunch
was developed in [12-17]. It was shown that optimum condition for wave
generating is $n_b=n_0/2$ ($n_b,$ $n_0$ are the density values of bunch and
plasma electrons, correspondingly). Important result of this theory is the
demonstration that, in the case of nonlinear wake fields a transformation
ratio $R=E_{ac}/E_{st}$ ($E_{ac}$ and $E_{st}$ are the extension of
correspondingly accelerating and decelerating electric fields) depends on
gamma-factor of accelerating bunch and may be significant without special
bunch shaping, as it occurs in the case of linear wake fields generated 
by a
rigid bunch. The conclusion of the results mentioned above is reaffirmed in
[18] for the assumption $\beta _0=v_0/c=1$, ($\gamma _0=(1-\beta
_0^2)^{-1/2}=\infty $) (where $v_0$ is the bunch velocity), that brings to
the incorrect expression for the maximum value of accelerating field $E_{ac}$
(when $n_b/n_0=1/2$ $E_{ac}=\infty $).

The influence of transverse sizes of the bunch on non-linear wake field
generated by short bunches ($d\ll \lambda _p,$ $d$ is the bunch length, $%
\lambda _p$ is the wave length) was considered in [19].

Non-linear theory of wake field generated by two- and three-dimensional
bunches for the general case is not yet developed.

This work contains the results and analysis of linear equation solution,
describing the interaction of the axial symmetrical homogenous bunch of
charged particles with plasma in the assumption of the plasma vorticity
absence (laminar flow), as well as at the strong external constant magnetic
field applied along the bunch motion, when the transverse movements of
plasma electrons are suppressed.

\section{Basic equations}

Vector equation, describing the excitation of nonlinear three-dimensional
wake fields by the rigid bunch of charged particles with electron density $%
n_b$ moving with constant velocity $v_0$ along the $z$ axis through the cold
plasma at equilibrium with density $n_0$ in hydrodynamic description and in
the assumption of absence of the plasma vorticity

\begin{equation}
{\rm rot}\left( {\bf \rho }-\frac e{mc^2}{\bf A}\right) =0,
\end{equation}
is given by the following formula [3]

\begin{equation}
\left( {\bf \nabla \nabla }-\nabla ^2+\frac 1{c^2}\frac{\partial ^2}{%
\partial t^2}\right) {\bf \rho +}\frac{{\bf \rho }}{\sqrt{1+\rho ^2}}\left[
\beta _0^2k_p^2\left( 1-\frac{n_b}{n_0}\right) +\frac 1c\frac \partial
{\partial t}{\bf \nabla \rho }+\nabla ^2\sqrt{1+\rho ^2}\right] +
\end{equation}

\[
+\frac 1c\frac \partial {\partial t}{\bf \nabla }\sqrt{1+\rho ^2}=-\beta
_0^3k_p^2\frac{n_b}{n_0}, 
\]
where ${\bf \rho }={\bf p}/mc$ is dimensionless momentum of plasma
electrons, ${\bf A}$-vector potential of electromagnetic field, $k_p=\omega
_p/v_0,$ $\omega _p=\sqrt{4\pi n_0e^2/m}$-is the plasma frequency of
electrons, $\beta _0=v_0/c,$ and $n_b$-arbitrary function of coordinates and
time.

The similar equation, describing interaction between laser pulse and the
plasma, was obtained in [20-22].

Let as consider axial-symmetrical bunch, when ${\bf \rho }$ depends only on
variable $r$ and $\widetilde{z}=z-v_0t$ (steady state). In this case 
vector $%
{\bf \rho }$ has only longitudinal $\rho _z$ and radial $\rho _r$
components, which do not depend on azimuthal angle $\varphi $, and $\rho
_\varphi =0.$

The system of equations for the component of momentum $\rho _z$ and $\rho _r$
has the following form:

\begin{eqnarray}
&&\frac{\partial ^2}{\partial \widetilde{z}^2}\left( \beta _0\rho _z-\sqrt{%
1+\rho ^2}\right) +\frac{\beta _0^2k_p^2\rho _z}{\beta _0\sqrt{1+\rho ^2}%
-\rho _z}+\beta _0^2k_p^2\frac{n_b}{n_0} \\
&=&\frac{\beta _0\rho _z-\sqrt{1+\rho ^2}}{\beta _0\sqrt{1+\rho ^2}-\rho _z}%
\frac \partial {\partial \widetilde{z}}\left( \frac 1r\frac \partial
{\partial r}\left( r\rho _r\right) \right) +\frac{\sqrt{1+\rho ^2}}{\beta _0%
\sqrt{1+\rho ^2}-\rho _z}\frac 1r\frac \partial {\partial r}\left( r\frac{%
\partial \rho _z}{\partial r}\right) -  \nonumber \\
&&-\frac{\rho _z}{\beta _0\sqrt{1+\rho ^2}-\rho _z}\frac 1r\frac \partial
{\partial r}\left( r\frac \partial {\partial r}\sqrt{1+\rho ^2}\right) , 
\nonumber
\end{eqnarray}

\begin{eqnarray}
&&\frac{\partial ^2}{\partial \widetilde{z}\partial r}\left( \beta _0\sqrt{%
1+\rho ^2}-\rho _z\right) +\frac 1{\gamma _0^2}\frac{\partial ^2\rho _r}{%
\partial \widetilde{z}^2} \\
&=&\frac{\rho _r}{\sqrt{1+\rho ^2}}\left\{ \beta _0^2k_p^2\left( 1-\frac{n_b%
}{n_0}\right) -\frac{\partial ^2}{\partial \widetilde{z}^2}\left( \beta
_0\rho _z-\sqrt{1+\rho ^2}\right) -\right.  \nonumber \\
&&\left. -\beta _0\frac \partial {\partial \widetilde{z}}\left( \frac
1r\frac \partial {\partial r}(r\rho _r)\right) +\frac 1r\frac \partial
{\partial r}\left( r\frac \partial {\partial r}\sqrt{1+\rho ^2}\right)
\right\} .  \nonumber
\end{eqnarray}

Inserting in (3) and (4) $\rho _r=0$ and $\rho _z=\rho _z(\widetilde{z})$
(dependence on $r$ is absent) we come to the one-dimensional nonlinear
equation, considered in [12-17]

\begin{equation}
\frac{\partial ^2}{\partial \widetilde{z}^2}\left( \beta _0\rho _z-\sqrt{%
1+\rho _z^2}\right) +\frac{\beta _0^2k_p^2\rho _z}{\beta _0\sqrt{1+\rho 
_z^2}%
-\rho _z}+\beta _0^2k_p^2\frac{n_b}{n_0}=0.
\end{equation}

Inserting in (4) $\rho _r=0$ and expressing the derivative $\partial \rho
_z/\partial \widetilde{z}$ through derivatives of $\rho _z$ we receive for
scalar potential $\varphi $ the following equation, describing interaction
of electron bunch with plasma in the presence of the external magnetic 
field 
${\bf B}_0$ $(0,0,B_0)$ [19]:

\begin{equation}
\frac 1r\frac \partial {\partial r}\left( r\frac{\partial \chi }{\partial r}%
\right) +\frac 1{\gamma _0^2}\frac{\partial ^2\chi }{\partial 
\widetilde{z}^2%
}+\beta _0^2k_p^2\left( 1-\frac{\beta _0\chi }{\sqrt{\chi ^2-1/\gamma _0^2}}%
\right) =\frac{\beta _0^2k_p^2}{\gamma _0^2}\frac{n_b}{n_0},
\end{equation}
where

\begin{equation}
\chi =1+\frac{e\varphi }{mc^2\gamma _0^2}=\sqrt{1+\rho _z^2}-\beta _0\rho _z.
\end{equation}

\section{Wakefield generating at vorticity absence}

Let us consider the problem of wake field generating by rigid cylindrical
bunch of radius $a$ and horizontal dimension $d$ with homogeneous
distribution of electrons of the bunch

\begin{equation}
n_b=\left\{ 
\begin{array}{l}
n_b,\quad{\ }0\leq r\leq a,\ 0\leq \widetilde{z}\leq d, \\ 
0,\quad{\ }a\leq r\leq b,\ \widetilde{z}\geq d,\ \widetilde{z}\leq 0,
\end{array}
\right.
\end{equation}
moving in conducting plasma waveguide of radius $b\geq a$.

Assuming in (3) and (4) $n_b/n_0\ll 1$ (linear approximation) and
linearizing the system on $\rho $ we shall obtain the following system of
equations, describing the process under consideration:

\begin{eqnarray}
\frac{\partial ^2\rho _r}{\partial \widetilde{z}\partial r}+\frac 1r\frac{%
\partial \rho _r}{\partial \widetilde{z}}-\frac{\partial ^2\rho _z}{\partial
r^2}-\frac 1r\frac{\partial \rho _z}{\partial r}+\beta _0^2\frac{\partial
^2\rho _z}{\partial \widetilde{z}^2}+\beta _0^2k_p^2\rho _z &=&-\beta
_0^3k_p^2\frac{n_b}{n_0}, \\
\frac{\partial ^2\rho _z}{\partial \widetilde{z}\partial r}-\frac 1{\gamma
_0^2}\frac{\partial ^2\rho _r}{\partial \widetilde{z}^2}+\beta _0^2k_p^2\rho
_r &=&0.  \nonumber
\end{eqnarray}

To define $\rho _z(r,\widetilde{z})$ and $\rho _r(r,\widetilde{z})$ let us
perform the Hankel transformation [23] of the equations system (9) on $r$ in
the finite limits ($0,$ $b$) and solving the obtained equations on $%
\widetilde{z}$ in the assumption of the continuity conditions of the
momentum components $\rho _z,$ $\rho _r$ and components of the electrical
field $E_z=\frac{mcv_0}e\frac{\partial \rho _z}{\partial \widetilde{z}},$ $%
E_r=\frac{mcv_0}e\frac{\partial \rho _r}{\partial \widetilde{z}}$ at the
front ($\widetilde{z}=d$) and rear ($\widetilde{z}=0$) bunch boundaries we
shall receive the following expression for the field components $E_z$ and $%
E_r$:

Inside the bunch ($0\leq r\leq a,$ $0\leq \widetilde{z}\leq d$)

\begin{eqnarray}
E_z^I &=&\frac{mv_0\omega _p}e\frac{n_b}{n_0}\sin k_p(d-\widetilde{z})\times
\\
&&\times \left\{ 1-(k_pa)\frac{I_0(k_pr)}{I_0(k_pb)}\left[
K_0(k_pb)I_1(k_pa)+I_0(k_pb)K_1(k_pa)\right] \right\} +E_{zH}^I,  \nonumber
\end{eqnarray}

\begin{eqnarray}
E_r^I &=&-\frac{mv_0\omega _p}e\frac{n_b}{n_0}(k_pa)\cos 
k_p(d-\widetilde{z})%
\frac{I_1(k_pr)}{I_0(k_pb)}\times \\
&&\times \left[ I_1(k_pa)K_0(k_pb)+I_0(k_pb)K_1(k_pa)\right] +E_{rH}^I; 
\nonumber
\end{eqnarray}

Over the bunch ($a\leq r\leq b,$ $0\leq \widetilde{z}\leq d$)

\begin{eqnarray}
E_z^{II} &=&\frac{mv_0\omega _p}e\frac{n_b}{n_0}\sin k_p(d-\widetilde{z}%
)(k_pa)\frac{I_1(k_pa)}{I_0(k_pb)}\times \\
&&\times \left[ K_0(k_pr)I_0(k_pb)-I_0(k_pr)K_0(k_pb)\right] 
+E_{zH}^{II}, 
\nonumber
\end{eqnarray}

\begin{eqnarray}
E_r^{II} &=&-\frac{mv_0\omega _p}e\frac{n_b}{n_0}(k_pa)\cos 
k_p(d-\widetilde{%
z})\frac{I_1(k_pa)}{I_0(k_pb)}\times \\
&&\times \left[ I_1(k_pr)K_0(k_pb)+I_0(k_pb)K_1(k_pr)\right] 
+E_{rH}^{II}, 
\nonumber
\end{eqnarray}
where

\begin{equation}
E_{zH}^I,\ E_{zH}^{II}=2\frac{mv_0\omega _p}e\frac{n_b}{n_0}\frac{(k_pa)}{%
b\gamma _0}\sum_{n=1}^\infty \frac{\mu _nJ_0\left( \frac{\mu _nr}b\right)
J_1\left( \frac{\mu _na}b\right) e^{-\gamma _0\kappa d/2}}{\kappa \left(
k_p^2b^2+\mu _n^2\right) J_1^2(\mu _n)}{\rm sh}\gamma _0\kappa \left( \frac
d2-\widetilde{z}\right) ,
\end{equation}

\begin{equation}
E_{rH}^I,\ E_{rH}^{II}=2\frac{mv_0\omega _p}e\frac{n_b}{n_0}%
(k_pa)\sum_{n=1}^\infty \frac{J_1\left( \frac{\mu _nr}b\right) J_1\left( 
\frac{\mu _na}b\right) e^{-\gamma _0\kappa d/2}}{\left( k_p^2b^2+\mu
_n^2\right) J_1^2(\mu _n)}{\rm ch}\gamma _0\kappa \left( \frac 
d2-\widetilde{%
z}\right)
\end{equation}
in the corresponding areas on $r$.

Before the bunch ($0\leq r\leq a\leq b,$ $d\leq \widetilde{z}\leq \infty $)
we have the following expression for the field components:

\begin{equation}
E_z^{III}=-2\frac{mv_0\omega _p}e\frac{n_b}{n_0}\frac{(k_pa)}{b\gamma _0}%
\sum_{n=1}^\infty \frac{\mu _nJ_0\left( \frac{\mu _nr}b\right) J_1\left( 
\frac{\mu _na}b\right) }{\kappa \left( k_p^2b^2+\mu _n^2\right) J_1^2(\mu 
_n)%
}e^{\gamma _0\kappa \left( \frac d2-\widetilde{z}\right) }{\rm sh}\left(
\gamma _0\kappa \frac d2\right) ,
\end{equation}

\begin{equation}
E_r^{III}=-2\frac{mv_0\omega _p}e\frac{n_b}{n_0}(k_pa)\sum_{n=1}^\infty 
\frac{J_1\left( \frac{\mu _nr}b\right) J_1\left( \frac{\mu _nb}a\right) }{%
\left( k_p^2b^2+\mu _n^2\right) J_1^2(\mu _n)}e^{\gamma _0\kappa \left(
\frac d2-\widetilde{z}\right) }{\rm sh}\left( \gamma _0\kappa \frac
d2\right) .
\end{equation}

After the bunch ($-\infty \leq \widetilde{z}\leq 0$) we have

\begin{eqnarray}
E_z^{IV} &=&\frac{mv_0\omega _p}e\frac{n_b}{n_0}\left[ \sin 
(k_p\widetilde{z}%
)+\sin k_p(d-\widetilde{z})\right] \times \\
&&\times \left\{ 1-(k_pa)\frac{I_0(k_pr)}{I_0(k_pb)}\left[
K_0(k_pb)I_1(k_pa)+I_0(k_pb)K_1(k_pa)\right] \right\} +E_{zH}^{IV}, 
\nonumber \\
0 &\leq &r\leq a,  \nonumber
\end{eqnarray}

\begin{eqnarray}
E_z^{IV} &=&\frac{mv_0\omega _p}e\frac{n_b}{n_0}\left[ \sin 
(k_p\widetilde{z}%
)+\sin k_p(d-\widetilde{z})\right] \times \\
&&\times (k_pa)\frac{I_1(k_pa)}{I_0(k_pb)}\left[
I_0(k_pb)K_0(k_pr)-K_0(k_pb)I_0(k_pr)\right] +E_{zH}^{IV},  \nonumber \\
a &\leq &r\leq b,  \nonumber
\end{eqnarray}

where

\begin{equation}
E_{zH}^{IV}=2\frac{mv_0\omega _p}e\frac{n_b}{n_0}\frac{(k_pa)}{b\gamma _0}%
\sum_{n=1}^\infty \frac{\mu _nJ_0\left( \frac{\mu _nr}b\right) J_1\left( 
\frac{\mu _na}b\right) }{\kappa \left( k_p^2b^2+\mu _n^2\right) J_1^2(\mu 
_n)%
}e^{-\gamma _0\kappa \left( \frac d2-\widetilde{z}\right) }{\rm sh}\left(
\gamma _0\kappa \frac d2\right) ,
\end{equation}

\begin{eqnarray}
E_r^{IV} &=&-2\frac{mv_0\omega _p}e\frac{n_b}{n_0}(k_pa)\sin \left( \frac{%
k_pd}2\right) \sin k_p\left( \widetilde{z}-\frac d2\right) \times \\
&&\times \left\{ 
\begin{array}{c}
\frac{I_1(k_pr)}{I_0(k_pb)}\left[
I_1(k_pa)K_0(k_pb)+I_0(k_pb)K_1(k_pa)\right] ,\quad \quad{at }0\leq r\leq a
\\ 
\frac{I_1(k_pa)}{I_0(k_pb)}\left[
K_0(k_pb)I_1(k_pr)+I_0(k_pb)K_1(k_pr)\right] ,\quad \quad{at }a\leq r\leq b
\end{array}
\right. -  \nonumber \\
&&-2\frac{mv_0\omega _p}e\frac{n_b}{n_0}(k_pa)\sum_{n=1}^\infty \frac{%
J_1\left( \mu _n\frac rb\right) J_1\left( \mu _n\frac ab\right) }{\left(
k_p^2b^2+\mu _n^2\right) J_1^2(\mu _n)}e^{-\gamma _0\kappa \left( \frac d2-%
\widetilde{z}\right) }{\rm sh}\left( \gamma _0\kappa \frac d2\right) . 
\nonumber
\end{eqnarray}

The components different from zero $B_\varphi $ of $E$-wave ($E_z$, 
$E_r$, $%
B_\varphi $) magnetic field is determining by the formula

\begin{equation}
B_\varphi =\frac{mc^2}e\left( \frac{\partial \rho _r}{\partial 
\widetilde{z}}%
-\frac{\partial \rho _z}{\partial r}\right)
\end{equation}
and is described by the following expressions

\begin{eqnarray}
B_\varphi ^{I,II} &=&-2\frac{mv_0\omega _p}e\beta _0(k_pa)\frac{n_b}{n_0}%
\sum_{n=1}^\infty \frac{J_1\left( \mu _n\frac rb\right) J_1\left( \mu
_n\frac ab\right) }{\kappa ^2b^2J_1^2(\mu _n)}\times \\
&&\times \left[ 1-e^{-\gamma _0\frac{\kappa d}2}{\rm ch}\gamma _0\kappa
\left( \frac d2-\widetilde{z}\right) \right] ,  \nonumber \\
\quad{at }0 &\leq &\widetilde{z}\leq d,\quad 0\leq r\leq b,  \nonumber
\end{eqnarray}

\begin{eqnarray}
B_\varphi ^{III} &=&-2\frac{mv_0\omega _p}e\beta _0(k_pa)\frac{n_b}{n_0}%
\sum_{n=1}^\infty \frac{J_1\left( \mu _n\frac rb\right) J_1\left( \mu
_n\frac ab\right) }{\kappa ^2b^2J_1^2(\mu _n)}e^{\gamma _0\kappa \left(
\frac d2-\widetilde{z}\right) }{\rm sh}\left( \gamma _0\frac{\kappa d}%
2\right) , \\
d &\leq &\widetilde{z}\leq \infty ,\quad 0\leq r\leq b,  \nonumber
\end{eqnarray}

\begin{eqnarray}
B_\varphi ^{IV} &=&-2\frac{mv_0\omega _p}e\beta _0(k_pa)\frac{n_b}{n_0}%
\sum_{n=1}^\infty \frac{J_1\left( \mu _n\frac rb\right) J_1\left( \mu
_n\frac ab\right) }{\kappa ^2b^2J_1^2(\mu _n)}e^{-\gamma _0\kappa \left(
\frac d2-\widetilde{z}\right) }{\rm sh}\left( \gamma _0\frac{\kappa d}%
2\right) , \\
-\infty &\leq &\widetilde{z}\leq 0,\quad 0\leq r\leq b.  \nonumber
\end{eqnarray}

In formulas (10)-(25) $J_0$, $J_1$ are the Bessel functions, and $I_0$, 
$I_1$%
, $K_0$, $K_1$-are the modified Bessel functions, $\kappa ^2=k_p^2\beta
_0^2+\mu _n^2/b^2$.

As it may be seen from the given expressions for field components $E_z$, $%
E_r $ they consist of periodic (wake field) and non-periodic (''Coulomb'')
parts. Before the bunch the field has only non-periodic part, which
exponentially falls of with the remote from the front boundary, so that part
may be neglected. After the bunch the ''Coulomb'' part also falls off
exponentially with the remote from the rear boundary $\widetilde{z}=0$ and
only periodic wake field remains. Magnetic field component $B_\varphi $ has
only non-periodic ''Coulomb'' part, which exponentially also decreases at
the remote from the bunch boundaries and it can be ignored.

In the range of $0\leq \widetilde{z}\leq d$, where magnetic field is not
small, the radial force $f_r=-eE_r+e\beta _0B_\varphi $ acting on the bunch
electrons, and this force in some ranges on $\widetilde{z}$ can compress the
bunch, focusing it. Before the bunch ($\widetilde{z}\geq d$) and in the
range of wake field ($\widetilde{z}\leq 0$) the component of the magnetic
field $B_\varphi $ is small and the radial force is $f_r\approx -eE_r$.

Below one can find an expression for the radial force $f_r$ acting on the
bunch ($0\leq \widetilde{z}\leq d$) at $\beta _0\approx 1$, $\gamma _0\gg 1$:

\begin{eqnarray}
f_r &=&-m\omega _pv_0\frac{n_b}{n_0}(k_pa)\left[ 1-\cos k_p\left( d-%
\widetilde{z}\right) \right] \times \\
&&\times \left\{ 
\begin{array}{l}
\frac{I_1(k_pr)}{I_0(k_pb)}\left[
I_1(k_pa)K_0(k_pb)+I_0(k_pb)K_1(k_pa)\right] ,\quad \quad r\leq a, \\ 
\frac{I_1(k_pa)}{I_0(k_pb)}\left[
K_0(k_pb)I_1(k_pr)+I_0(k_pb)K_1(k_pr)\right] ,\quad \ \quad a\leq r\leq b,
\end{array}
\right. +{\rm O}\left( \frac 1{\gamma _0^2}\right) ,  \nonumber
\end{eqnarray}
where ${\rm O}(\gamma _0^{-2})$ is the smaller defocusing part of the force.

Let us also give an expressions for the fields $E_z$ and $E_r$ inside and
after the bunch in the case when $a=b$ and in the case when $b\rightarrow
\infty $ (unlimited plasma):

\begin{eqnarray}
E_z^I &=&\frac{mv_0\omega _p}e\frac{n_b}{n_0}\sin k_p\left( d-\widetilde{z}%
\right) \left[ 1-\frac{I_0(k_pr)}{I_0(k_pa)}\right] +E_{zH}^I, \\
E_r^I &=&-\frac{mv_0\omega _p}e\frac{n_b}{n_0}\cos k_p\left( d-\widetilde{z}%
\right) \frac{I_1(k_pr)}{I_0(k_pa)}+E_{rH}^I,  \nonumber \\
E_z^{IV} &=&\frac{mv_0\omega _p}e\frac{n_b}{n_0}\left[ \sin k_p\left( d-%
\widetilde{z}\right) +\sin \left( k_p\widetilde{z}\right) \right] \left[ 1-%
\frac{I_0(k_pr)}{I_0(k_pa)}\right] +E_{zH}^{IV},  \nonumber \\
E_r^{IV} &=&-\frac{mv_0\omega _p}e\frac{n_b}{n_0}\left[ \cos k_p\left( d-%
\widetilde{z}\right) -\cos \left( k_p\widetilde{z}\right) \right] \frac{%
I_1(k_pr)}{I_0(k_pa)}-E_{rH}^{IV},  \nonumber \\
a &=&b,  \nonumber
\end{eqnarray}

where

\begin{eqnarray}
E_{zH}^I &=&2\frac{mv_0\omega _p}e\frac{n_b}{n_0}\frac{k_p}{\gamma _0}%
\sum_{n=1}^\infty \frac{\mu _nJ_0\left( \mu _n\frac ra\right) }{\kappa
J_1(\mu _n)\left( k_p^2a^2+\mu _n^2\right) }e^{-\gamma _0\frac{\kappa d}2}%
{\rm sh}\gamma _0\kappa \left( \frac d2-\widetilde{z}\right) , \\
E_{rH}^I &=&2\frac{mv_0\omega _p}e\frac{n_b}{n_0}(k_pa)\sum_{n=1}^\infty 
\frac{J_1\left( \mu _n\frac ra\right) }{\left( k_p^2a^2+\mu _n^2\right)
J_1(\mu _n)}e^{-\gamma _0\frac{\kappa d}2}{\rm ch}\gamma _0\kappa \left(
\frac d2-\widetilde{z}\right) ,  \nonumber
\end{eqnarray}
and $E_{zH}^{IV}$, $E_{rH}^{IV}\rightarrow 0$ at $\widetilde{z}<0$.

At the $a\rightarrow \infty $ $I_0(k_pa)\rightarrow \infty $ and the
expressions for $E_z^I$ and $E_z^{IV}$ coincide with the expressions for
one-dimensional bunch, while $E_r^I$ and $E_r^{IV}\rightarrow 0$. At $%
k_pa\ll 1$ ($k_pr<k_pa\ll 1$) periodic parts of the fields $E_z^{I,IV}$ 
and $%
E_r^{I,IV}$ are proportional to $(k_pa)^2$ and $k_pr$. Non-periodic parts
are small at $\gamma _0\gg 1$.

The expressions for $E_z$ and $E_r$ inside the bunch and in the wake field
at $b\rightarrow \infty $, $\gamma _0\gg 1$ go over into corresponding
expressions for the case with unlimited plasma. In this case at $k_pr\ll 
1$, 
$k_pa\ll 1$ ($r>a$ or $r<a$) the longitudinal fields inside and over the
bunch as well as in the wake are proportional to $(k_pa)^2$, while the
radial components of the fields $E_r^{I,IV}\sim r/2a$ at $r<a$ and $%
E_r^{I,IV}\sim a/2r$ at $r>a$, e.g. they increase with the remote from the
bunch centre inside the bunch and decrease with the remote from the bunch
boundaries inside it.

Let us define the plasma density for the mentioned four regions. From the
Poisson equation it follows that

\begin{equation}
n_e=-\frac 1{4\pi e}\left[ \frac{\partial E_r}{\partial 
r}+\frac{E_r}r+\frac{%
\partial E_z}{\partial \widetilde{z}}-4\pi e\left( n_0-n_b\right) \right] .
\end{equation}
In the regions outside the bunch in (29) we should suppose $n_b=0$. Using
the expressions (10)-(21) for the fields we shall find out the following
expressions for the plasma densities:

\begin{equation}
n_e^{I,II}=\left\{ 
\begin{array}{l}
n_0\left[ 1-\frac{n_b}{n_0}\left( 1-\cos k_p\left( d-\widetilde{z}\right)
\right) \right] ,\quad \quad r\leq a, \\ 
n_0,\quad \quad \quad \quad \quad \quad \quad \quad \quad \quad \quad \quad
\quad \quad a\leq r\leq b,
\end{array}
\right. 0\leq \widetilde{z}\leq d,
\end{equation}

\begin{equation}
n_e^{III}=n_0,\quad \quad{at }0\leq r\leq b,\quad d\leq \widetilde{z}<\infty
,
\end{equation}

\begin{equation}
n_e^{IV}=\left\{ 
\begin{array}{l}
n_0\left[ 1-\frac{n_b}{n_0}\left( \cos \left( k_p\widetilde{z}\right) -\cos
k_p\left( d-\widetilde{z}\right) \right) \right] ,\quad \quad r\leq a, \\ 
n_0,\quad \quad \quad \quad \quad \quad \quad \quad \quad \quad \quad \quad
\quad \quad \quad \quad \quad \quad a\leq r\leq b,
\end{array}
\right. -\infty <\widetilde{z}\leq 0.
\end{equation}

Thus, the density $n_e$ depends periodically on $\widetilde{z}$ inside and
after the electron bunch at $r\leq a$ and does not depend on $r$ and 
sizes $%
a $ and $b$.

For the regions outside the bunch $a\leq r\leq b$ and $0\leq z\leq \infty 
$, 
$-\infty \leq \widetilde{z}\leq 0$ $n_e^{II-IV}=n_0$ and the density $n_e$
undergoes change to the lateral surface of the bunch and in the wake at 
$r=a$%
. On the waveguide surface $r=b$, $n_e=n_0$.

\section{Wake field excitation under strong external magnetic field}

The linear equation for the potential $\varphi $, describing the interaction
of cylindrical electron bunch of radius $a$ and length $d$ with unlimited
cold plasma, follows from expressions (6), (7) in assumption $e\varphi
/mc^2\ll 1$, $n_b/n_0\ll 1$ and has the following form:

\begin{equation}
\frac{\partial ^2\varphi }{\partial r^2}+\frac 1r\frac{\partial \varphi }{%
\partial r}+\frac 1{\gamma _0^2}\frac{\partial ^2\varphi }{\partial 
\widetilde{z}^2}+\frac{k_p^2}{\gamma _0^2}\varphi =k_p^2v_0^2\frac me\frac{%
n_b}{n_0},
\end{equation}
where $n_b$ is given by the expression (8).

After making Hankel (Fourier-Bessel) transformation of the equation (33) 
on $%
r$ in boundless limits ($0$, $\infty $) under the condition that $\varphi (%
\widetilde{z},r\rightarrow \infty )=0$ we shall receive the following
equation

\begin{equation}
\frac{\partial ^2\overline{\varphi }(\alpha ,\widetilde{z})}{\partial 
\widetilde{z}^2}-\lambda ^2\overline{\varphi }(\alpha ,\widetilde{z})=h,
\end{equation}
where $\overline{\varphi }(\alpha ,\widetilde{z})=\int_0^\infty \varphi (r,%
\widetilde{z})J_0(\alpha r)rdr$, $\lambda ^2=\gamma _0^2(\alpha ^2-k^2)$, $%
k^2=k_p^2/\gamma _0^2$, $h=\frac mek_p^2v_0^2\frac{n_b}{n_0}\gamma _0^2\frac
a2J_1(\alpha a)$, $J_0$ and $J_1$-are the Bessel functions.

Before the bunch ($\widetilde{z}\geq d$), where $n_b=0$, $h=0$ the 
potential 
$\varphi (r,\widetilde{z})$ is assumed as equal to zero (we ignore the
''Coulomb'' field, see chapter 3).

Solving the equation under assumption of continuity of the potential $%
\overline{\varphi }(\alpha ,\widetilde{z})$ on the front ($\widetilde{z}=d$)
and rear ($\widetilde{z}=0$) boundaries of the bunch, we shall come to the
following expressions for $\varphi (r,\widetilde{z})$ in the ranges inside
and over the bunch ($0\leq \widetilde{z}\leq d$, $0\leq r\leq a$, $a\leq
r<\infty $)

\begin{equation}
\varphi _1=-\frac mek_p^2v_0^2\frac{n_b}{n_0}a{\rm Re}\int_0^\infty \frac{%
J_0(\alpha r)J_1(\alpha a)}{\alpha ^2-k^2}\left[ 1-e^{-\lambda 
(d-\widetilde{%
z})}\right] d\alpha ,
\end{equation}
where $k$ has the positive imaginary part ${\rm Im}k=k^{^{\prime }}>0$, $%
\lambda =\gamma _0\sqrt{\alpha ^2-k^2}$ at $\alpha >k$, $\lambda 
=-i\gamma _0%
\sqrt{k^2-\alpha ^2}$ at $\alpha <k$.

After the bunch ($\widetilde{z}\leq 0$, $0\leq r<\infty $) the potential $%
\varphi $ has the following form:

\begin{equation}
\varphi _2=-\frac mek_p^2v_0^2\frac{n_b}{n_0}a{\rm Re}\int_0^\infty \frac{%
J_0(\alpha r)J_1(\alpha a)}{\alpha ^2-k^2}\left[ e^{\lambda \widetilde{z}%
}-e^{\lambda (\widetilde{z}-d)}\right] d\alpha .
\end{equation}
Because the ''Coulomb'' components ($\alpha ^2>k^2$) are small one can
ignore them and bring out the expression of square brackets from the
integral in the point $\alpha =0$, where it makes the main input into the
integral. In this case the expressions (35), (36) significantly simplify:

\begin{equation}
\varphi _1=-\frac mek_p^2v_0^2\frac{n_b}{n_0}a{\rm Re}\left[ 1-e^{ik_p(d-%
\widetilde{z})}\right] \int_0^\infty \frac{J_0(\alpha r)J_1(\alpha a)}{%
\alpha ^2-k^2}d\alpha ,
\end{equation}

\begin{equation}
\varphi _2=-\frac mek_p^2v_0^2\frac{n_b}{n_0}a{\rm Re}\left[ e^{-ik_p%
\widetilde{z}}-e^{-ik_p(\widetilde{z}-d)}\right] \int_0^\infty \frac{%
J_0(\alpha r)J_1(\alpha a)}{\alpha ^2-k^2}d\alpha .
\end{equation}

The components of the fields are determined from the expressions:

\begin{equation}
E_z=-\frac 1{\gamma _0^2}\frac{\partial \varphi }{\partial \widetilde{z}}%
,\quad E_r=-\frac{\partial \varphi }{\partial r},\quad B_\varphi =\beta
_0E_r.
\end{equation}
Inside the bunch ($0\leq r\leq a$, $0\leq \widetilde{z}\leq d$) we have:

\begin{eqnarray}
E_z^I(r,\widetilde{z}) &=&\frac{mv_0\omega _p}e\frac{n_b}{n_0}\left\{ \sin
k_p\left( d-\widetilde{z}\right) \left[ 1+\frac \pi
2(ka)J_0(kr)Y_1(ka)\right] -\right. \\
&&\left. -\frac \pi 2(ka)\cos k_p\left( d-\widetilde{z}\right)
J_0(kr)J_1(ka)\right\} ,  \nonumber
\end{eqnarray}

\begin{eqnarray}
E_r^I(r,\widetilde{z}) &=&\frac{mv_0\omega _p}e\frac 1{\gamma _0}\frac{n_b}{%
n_0}\left\{ \frac \pi 2\left[ 1-\cos k_p\left( d-\widetilde{z}\right)
\right] (ka)J_1(kr)Y_1(ka)-\right. \\
&&\left. -\frac \pi 2(ka)\sin k_p\left( d-\widetilde{z}\right)
J_1(kr)J_1(ka)\right\} ,  \nonumber
\end{eqnarray}
where $k=k_p/\gamma _0$.

Over the bunch ($a\leq r<\infty $, $0\leq \widetilde{z}\leq d$) the field
components are given the following expressions:

\begin{eqnarray}
E_z^{II}(r,\widetilde{z}) &=&\frac{mv_0\omega _p}e\frac{n_b}{n_0}\left\{
\frac \pi 2(ka)\sin k_p\left( d-\widetilde{z}\right) J_1(ka)Y_0(kr)-\right.
\\
&&\left. -\frac \pi 2(ka)\cos k_p\left( d-\widetilde{z}\right)
J_0(kr)J_1(ka)\right\} ,  \nonumber
\end{eqnarray}

\begin{eqnarray}
E_r^{II}(r,\widetilde{z}) &=&\frac{mv_0\omega _p}e\frac 1{\gamma 
_0}\frac{n_b%
}{n_0}\left\{ \frac \pi 2(ka)\left[ 1-\cos k_p\left( d-\widetilde{z}\right)
\right] J_1(ka)Y_1(kr)-\right. \\
&&\left. -\frac \pi 2(ka)\sin k_p\left( d-\widetilde{z}\right)
J_1(kr)J_1(ka)\right\} .  \nonumber
\end{eqnarray}

The wake field is determined by the following expressions:

\begin{eqnarray}
E_z^{IV}(r,\widetilde{z}) &=&\frac{mv_0\omega _p}e\frac{n_b}{n_0}\left\{
\left[ \sin \left( k_p\widetilde{z}\right) -\sin k_p\left( \widetilde{z}%
-d\right) \right] \left[ 1+\frac \pi 2(ka)J_0(kr)Y_1(ka)\right] +\right. \\
&&\left. +\left[ \cos \left( k_p\widetilde{z}\right) -\cos k_p\left( 
\widetilde{z}-d\right) \right] \frac \pi 2(ka)J_0(kr)J_1(ka)\right\} , 
\nonumber \\
0 &\leq &r\leq a,  \nonumber
\end{eqnarray}

\begin{eqnarray}
E_z^{IV}(r,\widetilde{z}) &=&\frac{mv_0\omega _p}e\frac{n_b}{n_0}\left\{
\left[ \sin \left( k_p\widetilde{z}\right) -\sin k_p\left( \widetilde{z}%
-d\right) \right] \frac \pi 2(ka)J_1(ka)Y_0(kr)+\right. \\
&&\left. +\frac \pi 2(ka)\left[ \cos \left( k_p\widetilde{z}\right) -\cos
k_p\left( \widetilde{z}-d\right) \right] J_1(ka)J_0(kr)\right\} ,  \nonumber
\\
a &\leq &r<\infty ,  \nonumber
\end{eqnarray}

\begin{eqnarray}
E_r^{IV}(r,\widetilde{z}) &=&\frac{mv_0\omega _p}e\frac 1{\gamma 
_0}\frac{n_b%
}{n_0}\left\{ \left[ \cos \left( k_p\widetilde{z}\right) -\cos k_p\left( 
\widetilde{z}-d\right) \right] \frac \pi 2(ka)J_1(kr)Y_1(ka)-\right. \\
&&\left. -\left[ \sin \left( k_p\widetilde{z}\right) -\sin k_p\left( 
\widetilde{z}-d\right) \right] \frac \pi 2(ka)J_1(kr)J_1(ka)\right\} , 
\nonumber \\
0 &\leq &r\leq a,  \nonumber
\end{eqnarray}

\begin{eqnarray}
E_r^{IV}(r,\widetilde{z}) &=&\frac{mv_0\omega _p}e\frac 1{\gamma 
_0}\frac{n_b%
}{n_0}\left\{ \left[ \cos \left( k_p\widetilde{z}\right) -\cos k_p\left( 
\widetilde{z}-d\right) \right] \frac \pi 2(ka)J_1(ka)Y_1(kr)-\right. \\
&&\left. -\left[ \sin \left( k_p\widetilde{z}\right) -\sin k_p\left( 
\widetilde{z}-d\right) \right] \frac \pi 2(ka)J_1(kr)J_1(ka)\right\} , 
\nonumber \\
a &\leq &r<\infty .  \nonumber
\end{eqnarray}

The components of the magnetic field $B_\varphi $ is determined from the
expression $B_\varphi =\beta _0E_r$.

The comparison of the expressions for the components of the fields
(10)-(25), generated by the electron bunch in plasma without magnetic field,
with the corresponding expressions (40)-(47) with the constant longitudinal
strong magnetic field ${\bf B}_0$ shows, that in the last case the character
of the fields qualitatively varies from the case of plasma without the
field. On the one hand, the dependence on the transversal coordinates $r$
and $a$ is determined by the Bessel functions $Y$ and $J$, having the
oscillating character, and on the other hand, there is more distinctly
expressed dependence on $\gamma $-factor of the bunch. Besides, magnetic
wake field $B_\varphi $ is equal to zero in plasma without the field and
defers from zero in the case of plasma with $B_0\neq 0$.

\begin{center}
{\bf Acknowledgment}
\end{center}

The work was supported by the $ISTC$ Grant $A-013$.

\begin{center}
{\bf References}
\end{center}

\begin{enumerate}
\item[{[1]}]  E. Esarey, P. Sprangle, J. Krall, IEEE Trans. Plasma Sci. 
{\bf %
24,} 252 (1996).

\item[{[2]}]  Ya. Feinberg, Fizika Plazmi {\bf 23,} 275 (1997) (in Russian).

\item[{[3]}]  A.Ts. Amatuni, S.S. Elbakian, A.G. Khachatryan, E.V.
Sekhpossian, Part. Acc. {\bf 51,} 1 (1995).

\item[{[4]}]  P. Chen, Part. Acc. {\bf 20,} 171 (1987).

\item[{[5]}]  P. Chen, J.M. Dawson, R.W. Huff, T. Katsouleas, Phys. Rev.
Lett. {\bf 54,} 693 (1985).

\item[{[6]}]  R.D. Ruth, A.W. Chao, P.L. Morton, P.B. Wilson, Part. Acc. 
{\bf 17,} 171 (1985).

\item[{[7]}]  R. Keinigs, M.E. Jones, Phys. Fluids {\bf 30}, 252 (1987).

\item[{[8]}]  V. Balakirev, Preprint 87-40, Kharkov: KhFTI, 1987 (in
Russian).

\item[{[9]}]  A. Amatuni, E. Sekhpossian, A. Khachatryan, S. Elbakian,
Plasma Phys. Rep {\bf 21,} 945 (1995).

\item[{[10]}]  A. Khachatryan, A. Amatuni, E. Sekhpossian, S. Elbakian,
Plasma Phys. Rep {\bf 22,} 576 (1996).

\item[{[11]}]  Ya. Feinberg, N. Ayzatskij, V. Balakirev et al., Proc. of XV
Int. Workshop on Charged Part. Linear Acc., 1997, Alushta, p. 16 (in
Russian).

\item[{[12]}]  A. Amatuni, M. Magomedov, E. Sekhpossian, S. Elbakian, Fizika
Plasmi {\bf 5,} 85 (1979) (in Russian).

\item[{[13]}]  A. Amatuni, E. Sekhpossian, S. Elbakian, Fizika Plasmi 
{\bf %
12,} 1145 (1986) (in Russian).

\item[{[14]}]  A.Ts. Amatuni, S.S. Elbakian, E.V. Sekhpossian, Prep.
YerPhI-935(86)-86, Yerevan, 1986.

\item[{[15]}]  A. Amatuni, E. Sekhpossian, S. Elbakian, Proc. of XIII Int.
Conf. on High Energy Part. Acc., 1987, Novosibirsk, Nauka Publ., v. 1, p.
175 (in Russian).

\item[{[16]}]  A.Ts. Amatuni, S.S. Elbakian, E.V. Sekhpossian, R.O.
Abramian, Part. Acc. {\bf 41}, 153 (1993); Preprint EFI-1365(60), Yerevan,
1991.

\item[{[17]}]  A.Ts. Amatuni, E.V. Sekhpossian, A.G. Khachatryan, S.S.
Elbakian, Izvestiya Akademii Nauk Armenii, Fizika {\bf 28}, 8 (1993): Soviet
J. of Cont. Phys., Allerton Press Inc. {\bf 28}, 7 (1994).

\item[{[18]}]  J.B. Rosenzweig, Phys. Rev. Lett. {\bf 88}, 555 (1987).

\item[{[19]}]  A.Ts. Amatuni, E.V. Sekhpossian, S.S. Elbakian, Izvestiya
Akademii Nauk Armenii, Fizika {\bf 25}, 308 (1990): Soviet J. of Cont.
Phys., Allerton Press Inc. {\bf 25}, 1 (1990).

\item[{[20]}]  W.B. Mori, T. Katsouleas, Proc. of EPAC-90, Nice-Paris, 1990,
v. 1, p. 603.

\item[{[21]}]  J. Krall, E. Esarey, P. Sprangle, G. Joice, Phys. Plasmas 
{\bf 1}, 1738 (1994).

\item[{[22]}]  N. Andrejev, L. Gorbunov, R. Ramazashvili, Fizika Plasmi 
{\bf %
23}, 303 (1997) (in Russian).

\item[{[23]}]  I. Snedon, Preobrazovanija Fourier, Moscow, Inostr. Lit.
Publ., 1955 (in Russian).
\end{enumerate}

\end{document}